\documentclass[letter]{IEEEtran}
\usepackage{amssymb}
\usepackage{amsmath,amsfonts,amssymb}
\usepackage[dvips]{graphicx}
\usepackage{verbatim}
\usepackage{slashbox}
\usepackage{bm}
\usepackage{algorithm} 
\usepackage{algorithmic} 
\usepackage{multirow} 
\usepackage{amsmath}
\usepackage{xcolor}
\usepackage{cite}

\usepackage[bookmarks=false]{}

\IEEEoverridecommandlockouts

\begin{document}

\title{Iterative Joint Beamforming Training with Constant-Amplitude Phased Arrays in Millimeter-Wave Communications}

\author{
Zhenyu Xiao,~\IEEEmembership{Member,~IEEE},
Lin Bai,~\IEEEmembership{Member,~IEEE},
and Jinho Choi,~\IEEEmembership{Senior Member,~IEEE}
\thanks{This work was partially supported by the National Natural Science Foundation of China under Grant Nos. 61201189, 91338106, and 61231011, as well as GIST College¡¯s 2013 GUP Research Fund.}
\thanks{Z. Xiao and L. Bai are with the School of
Electronic and Information Engineering, Beihang University, Beijing 100191, P.R. China.}
\thanks{J. Choi is with the School of Information and Communications, Gwangju Institute of Science and Technology (GIST), Korea.}
\thanks{Corresponding author is Dr. Z. Xiao (Email: xiaozy06@gmail.com).}
}

\maketitle

\begin{abstract}
In millimeter-wave communications (MMWC), in order to compensate for high propagation attenuation, phased arrays are favored to achieve array gain by beamforming, where transmitting and receiving antenna arrays need to be jointly trained to obtain appropriate antenna weight vectors (AWVs). Since the amplitude of each element of the AWV is usually constraint constant to simplify the design of phased arrays in MMWC, the existing singular vector based beamforming training scheme cannot be used for such devices. Thus, in this letter, a steering vector based iterative beamforming training scheme, which exploits the directional feature of MMWC channels, is proposed for devices with constant-amplitude phased arrays. Performance evaluations show that the proposed scheme achieves a fast convergence rate as well as a near optimal array gain. 
\end{abstract}


\vspace{-0.1 in}
\section{Introduction}
\IEEEPARstart{W}{hile arousing} increasing attentions in both academia and industry due to abundant frequency spectrum \cite{Xia_2011_60GHz_Tech,xiaozhenyu2013div}, millimeter-wave communications (MMWC) face the challenge of high propagation attenuation caused by the high frequency. To remedy this, phased arrays can be adopted at both the source and destination devices to exploit array gains and compensate for the propagation loss \cite{Xia_2011_60GHz_Tech,xiaozhenyu2013div}.

To achieve a sufficiently high array gain, the antenna weight vectors (AWVs)
at both ends need to be appropriately set prior to signal transmissions, which is the \emph{joint} Tx/Rx beamforming.
If the channel state information (CSI) is
available at both ends, the optimum AWVs can be directly
found under well-known
performance criteria, e.g., maximizing receiving signal-to-noise
ratio (SNR) \cite{Xia_2011_60GHz_Tech,xiaozhenyu2013div}.
Unfortunately, since the number of antennas is generally large,
the channel estimation becomes time-consuming in MMWC.
In addition, the computational complexity
is high, because the matrix decomposition, e.g., the singular vector
decomposition (SVD), is generally required. Owing to this, the beamforming
training approach, which has a lower complexity,
becomes more attractive to find the
AWVs \cite{wang_2009_beam_codebook,xia_2008_prac_SDMA,tang_2005}.
Generally, there are two types of joint Tx/Rx beamforming training schemes. One
is switched beamforming training based on a fixed codebook
\cite{wang_2009_beam_codebook}. The codebook contains
a number of predefined AWVs. During beamforming training, the AWVs
at both ends are examined according to a certain order, and the pair
that achieves the largest SNR is selected.
The other one is adaptive beamforming
training \cite{xia_2008_prac_SDMA,tang_2005}, which does not need
a fixed codebook. The desired AWVs at both ends are found
via real-time joint iterative training.
It is clear that the switched beamforming training is simpler,
while the adaptive one is more flexible.

Most adaptive beamforming training schemes
adopt the same state-of-the-art approach, i.e.,
finding the best singular vector via iterative training
without \emph{a priori} CSI at both ends \cite{xia_2008_prac_SDMA,tang_2005}.
This singular vector based training scheme (SGV) requires that both the amplitudes and phases of the AWV are adjustable. On the other hand, in MMWC, phased arrays are usually implemented with the approach that only phases of the antenna branches are adjustable; the amplitudes are set constant to simplify the design and reduce the power consumption of phased arrays \cite{wang_2009_beam_codebook,valdes2010fully,cohen2010thirty}. In fact, even in general multiple-input multiple-output (MIMO) systems \cite{bai2013lattice}, antenna branches with
constant amplitude (CA) are also an optimization objective to reduce implementation complexity \cite{zheng2007mimo,lee2009generalized}. In such a case, SGV becomes infeasible due to the CA phased array. The schemes proposed in \cite{zheng2007mimo,lee2009generalized} cannot be used here either, because these schemes are designed for transmitting beamforming with full or quantized \emph{a priori} CSI at only the source device, but for MMWC with CA phased arrays, joint beamforming is required without \emph{a priori} CSI at both the source and destination devices.

In this letter, a steering vector based joint beamforming training scheme (STV),
which exploits the directional feature of MMWC channels, is proposed.
Performance comparisons show that for line-of-sight (LOS) channels,
both STV and SGV have fast convergence rates and achieve the optimal
array gain. On the other hand,
for non-LOS (NLOS) channels, STV achieves a faster
convergence rate at the cost of a slightly lower array gain than SGV,
which can also achieve the optimal array gain. In conclusion,
STV achieves a fast convergence rate and a near optimal array gain
under both LOS and NLOS channels, which highlights its applicability
in practice.


\vspace{-0.1 in}

\section{System and Channel Models}


Without loss of generality, we consider a MMWC system with
half-wave spaced uniform linear arrays (ULAs) of $M$ and $N$ elements
at the source and destination devices, respectively.
The ULAs are phased arrays where only the phase
can be controlled. A single RF chain is tied to the ULA at
each of the source and destination devices.
According to the measurement results of channels
for MMWC in \cite{Xia_2011_60GHz_Tech,maltsev_2010},
only reflection contributes to generating multipath components (MPCs),
while scattering and diffraction effects are negligible due to the extremely
small wave length of MMWC. Thus, the MPCs in MMWC have a directional
feature, i.e., different MPCs have different physical transmitting
steering angles $\phi_{{\rm{t}}\ell}$ and receiving steering
angles $\phi_{{\rm{r}}\ell}$. Consequently,
the channel model is expressed as \cite{Park_2012_beam_diversity,TseFundaWC}
\vspace{-0.0 in}
\begin{equation} \label{eq_ch_nature}
\begin{aligned}
 {\bf{H}} &= \sqrt {{N}{M}} \sum\nolimits_{\ell  = 0}^{L - 1} {{\lambda _\ell }{{\bf{g}}_\ell }{\bf{h}}_\ell ^{\rm{H}}},
 \end{aligned}
\end{equation}
where $L$ is the number of multipath components,
$(\cdot)^{\rm{H}}$ is the conjugate transpose operation,
$\lambda_\ell$ are the channel coefficients,
and ${{\bf{g}}_\ell }$ and ${{\bf{h}}_\ell }$ are
the receiving and transmitting steering vectors
\cite{Park_2012_beam_diversity,TseFundaWC} that are given by ${{\bf{g}}_\ell } =\{{{\rm{e}}^{j\pi (n-1){\Omega _{{\rm{r}}\ell }}}}/\sqrt{N}\}_{n=1,2,...,N}$ and ${{\bf{h}}_\ell } =\{{{\rm{e}}^{j\pi (m-1){\Omega _{{\rm{t}}\ell }}}}/\sqrt{M}\}_{m=1,2,...,M}$,
respectively. Note that $\Omega_{{\rm{t}}\ell}$ and $\Omega_{{\rm{r}}\ell}$ represent the transmitting and receiving cosine angles of the $\ell$th MPC, respectively \cite{TseFundaWC},
i.e., $\Omega_{{\rm{t}}\ell}=\cos(\phi_{{\rm{t}}\ell})$ and $\Omega_{{\rm{r}}\ell}=\cos(\phi_{{\rm{r}}\ell})$.

Given the transmitting AWV ${\bf{t}}$ and the receiving AWV ${\bf{r}}$,
where $\|{\bf{t}}\|=\|{\bf{r}}\|=1$, the received signal $y$ is given by
$
y = {{\bf{r}}^{\rm{H}}}{\bf{Ht}}s + {{\bf{r}}^{\rm{H}}}{\bf{n}},
$
where $s$ is the transmitted symbol, ${\bf{n}}$ is the noise vector.
The target of beamforming training is to find appropriate transmitting
and receiving AWVs to obtain a high receiving SNR, which is given by
$
\gamma  = {{|{{\bf{r}}^{\rm{H}}}{\bf{Ht}}{|^2}}}/{{{\sigma ^2}}},
$
where $\sigma^2$ is the noise power.

\vspace{-0.1 in}
\section{Singular-Vector Based Scheme}
\begin{algorithm}[t]\caption{The SGV Scheme}\label{alg:SGV}
\begin{algorithmic}
\STATE \textbf{1) Initialize:}
\begin{quote}
Pick an initial transmitting AWV ${\bf{t}}$ at the source device. This AWV may be chosen either randomly or with the approach specified in Section V.
\end{quote}

\STATE \textbf{2) Iteration:}
Iterate the following process $\varepsilon$ times, and then stop.

\begin{quote}
Keep sending with the same transmitting AWV ${\bf{t}}$ at the source device over $N$ slots. Meanwhile, use identity matrix ${\bf{I}}_N$ as the receiving AWVs at the destination device, i.e., the $n$th column of ${\bf{I}}_N$ as the receiving AWV at the $n$th slot. Consequently, we receiving a vector
${\bf{r}} = {\bf{I}}_N^{\rm{H}}{\bf{Ht}} + {\bf{I}}_N^{\rm{H}}{{\bf{n}}_{\rm{r}}}={\bf{Ht}}+{{\bf{n}}_{\rm{r}}}$,
where ${\bf{n}}_{\rm{r}}$ is the noise vector. Normalize ${\bf{r}}$.

 Keep sending with the same transmitting AWV ${\bf{r}}$ at the destination device over $M$ slots. Meanwhile, use identity matrix ${\bf{I}}_M$ as the receiving AWVs at the source device. Consequently, we receiving a new vector
${\bf{t}} = {\bf{I}}_M^{\rm{H}}{{\bf{H}}^{\rm{H}}}{\bf{r}} + {\bf{I}}_M^{\rm{H}}{{\bf{n}}_{\rm{t}}}={{\bf{H}}^{\rm{H}}}{\bf{r}}+{{\bf{n}}_{\rm{t}}}$,
where ${\bf{n}}_{\rm{t}}$ is the noise vector. Normalize ${\bf{t}}$.
\end{quote}

\STATE \textbf{3) Result:}
\begin{quote}
${\bf{t}}$ is the AWV at the source device, and ${\bf{r}}$ is the AWV at the destination device.
\end{quote}
\end{algorithmic}
\end{algorithm}

Let us introduce the SGV scheme first. It is known that the optimal AWVs to maximize $\gamma$ is the principal singular vectors of the channel matrix ${\bf{H}}$ \cite{xia_2008_prac_SDMA,tang_2005}\footnote{The SVD on ${\bf{H}}$ gives a set of orthogonal transmitting and receiving AWV pairs, as well as the energies projected to these AWV pairs.}. Denote the SVD of ${\bf{H}}$ as
\vspace{-0.0 in}
\begin{equation} \label{eq:ch_svd}
{\bf{H}} = {\bf{U}}\Sigma {{\bf{V}}^{\rm{H}}} = \sum\nolimits_{k = 1}^K {{\rho _k}{{\bf{u}}_k}{\bf{v}}_k^{\rm{H}}},
\end{equation}
where ${\bf{U}}$ and ${\bf{V}}$ are unitary matrices with column vectors (singular vectors) ${\bf{u}}_k$ and ${\bf{v}}_k$, respectively, $\Sigma$ is an $N\times M$ rectangular diagonal matrix with nonnegative real values
${\rho_k}$ on the diagonal, i.e., ${\rho _1} \ge {\rho _2} \ge ... \ge {\rho _K} \ge 0$ and $K=\min(\{M,N\})$. The optimal AWVs are ${\bf{t}}={\bf{v}}_1$ and ${\bf{r}}={\bf{u}}_1$.

In the common case that ${\bf{H}}$ is unavailable, iterative beamforming training can be adopted to find the optimal AWVs. According to \cite{xia_2008_prac_SDMA,tang_2005},
$
{{\bf{H}}^{2m}} \buildrel \Delta \over = ({{\bf{H}}^{\rm{H}}}{\bf{H}})^m = \sum_{k = 1}^K {\rho _k^{2m}{{\bf{v}}_k}{\bf{v}}_k^{\rm{H}}}
$, which can be obtained by $m$ iterative trainings utilizing the reciprocal feature of the channel. When $m$ is large, ${{\bf{H}}^{2m}}\approx \rho_1^{2m}{{\bf{v}}_1}{\bf{v}}_1^{\rm{H}}$. Thus, the optimal transmitting and receiving AWVs can be obtained by normalizing ${{\bf{H}}^{2m}}{\bf{t}}$ and ${\bf{H}} \times {{\bf{H}}^{2m}}{\bf{t}}$, respectively.

The SGV scheme is described in Algorithm \ref{alg:SGV}. The iteration number $\varepsilon$ depends on practical channel response, which will be shown in Section V.
It is clear that although the SGV scheme is effective,
it is required that both the amplitudes and phases
of AWV are adjustable,
which cannot be satisfied when CA phased arrays are used, where only phases are adjustable.


 \vspace{-0.1 in}
\section{Steering-Vector Based Scheme}
In fact, the SGV scheme is a general one suitable for an arbitrary channel ${\bf{H}}$. It does not use the specific feature of MMWC channels.
In MMWC, the channel has a directional feature, i.e., ${\bf{H}}$ can be naturally expressed as in (\ref{eq_ch_nature}), which is similar to the expression in (\ref{eq:ch_svd}). The difference is that in (\ref{eq_ch_nature}),
the vectors $\{{\bf{g}}_\ell\}$ and $\{{\bf{h}}_\ell\}$ are CA steering vectors, not orthogonal bases, but in (\ref{eq:ch_svd}), $\{{\bf{u}}_k\}$ and $\{{\bf{v}}_k\}$ are strict non-CA orthogonal bases. Nevertheless, according to \cite{TseFundaWC}, $|{\bf{g}}_m^{\rm{H}}{\bf{g}}_n|$ and $|{\bf{h}}_m^{\rm{H}}{\bf{h}}_n|$ are approximately equal to zero given that $|\Omega_{{\rm{r}}m}-\Omega_{{\rm{r}}n}|\geq 1/N$ and $|\Omega_{{\rm{t}}m}-\Omega_{{\rm{t}}n}|\geq 1/M$, respectively, i.e., the receiving and transmitting angles can be resolved by the arrays, which is the common case in MMWC. Consequently, as a suboptimal approach, the steering vectors
of the \emph{strongest} MPC
can be adopted
as the transmitting and receiving AWVs at the source and destination devices,
which leads to
the proposed STV scheme.
The advantage of STV is that the elements of
the steering vector have a constant envelope, which is suitable
for the devices with CA phased arrays. Moreover, although the transmitting
and receiving angles are required to be resolved by the arrays in the following analysis, the STV scheme can work even when there exists angles that cannot be resolved,
because two or more MPCs associated with sufficiently close angles
that cannot be resolved actually
build a single equivalent MPC.

Assuming that ${\bf{H}}$ is available in advance,
the background of STV is presented as follows.
Using the directional feature of MMWC channels,
we have ${{\bf{H}}^{2m}} \approx \sum_{\ell = 1}^L {|\sqrt{MN}\lambda _\ell|^{2m}{{\bf{h}}_\ell}{\bf{h}}_\ell^{\rm{H}}}$,
for a positive integer $m$.
Suppose the $k$th MPC is the strongest one. For $\ell\neq k$, $|\lambda _\ell|^{2m}/|\lambda _k|^{2m}$ exponentially decreases.
This means that the contribution to the matrix
product ${{\bf{H}}^{2m}}$ from the the other $L-1$ MPCs exponentially
decreases, compared with the strongest one.
Therefore, we have $\mathop {\lim }\limits_{m \to \infty } {{\bf{H}}^{2m}} = |\sqrt{MN}\lambda _k|^{2m}{{\bf{h}}_k}{\bf{h}}_k^{\rm{H}}$.
Thus, for given a sufficiently large $m$ and an
arbitrary initial transmitting AWV ${\bf{t}}$, we have
\[
{{\bf{H}}^{2m}}{\bf{t}} = |\sqrt{MN}\lambda _k|^{2m}{{\bf{h}}_k}{\bf{h}}_k^{\rm{H}}{\bf{t}} = \left(|\sqrt{MN}\lambda _k|^{2m}{\bf{h}}_k^{\rm{H}}{\bf{t}}\right){{\bf{h}}_k},
\]
 which is ${{\bf{h}}_k}$ multiplied by a complex coefficient. It is noted that ${{\bf{h}}_k}$ is a constant-envelope steering vector. Hence, the desired transmitting AWV can be obtained by
the signature estimation\footnote{There are other approaches to obtain the desired AWV here. The presented one is a simple one in implementation.}
where $
{\bf{e}}_{\rm{t}}=\exp({\rm{j}}\measuredangle({{\bf{H}}^{2m}}{\bf{t}}))/{\sqrt{M}}$ is to be estimated.
Here, $\measuredangle({\bf{x}})$ represents the angle vector of ${\bf{x}}$ in radian. In fact, the signature estimation can be
carried out by the entry-wise normalization on ${{\bf{H}}^{2m}}{\bf{t}}$.

In addition, we have
\begin{equation}
\begin{aligned}
{\bf{H}} \times {{\bf{H}}^{2m}}{\bf{t}}= \left( \lambda _k\sqrt{MN}{|\lambda _k\sqrt{MN}|^{2m }{\bf{h}}_k^{\rm{H}}{\bf{t}}} \right){{\bf{g}}_k}.
\end{aligned}
\end{equation}
Thus, the desired receiving AWV can be obtained by
the signature estimation of $
{\bf{e}}_{\rm{r}}=\exp({\rm{j}}\measuredangle({\bf{H}} \times {{\bf{H}}^{2m}}{\bf{t}}))/{\sqrt{N}}$.

\begin{figure}[t]
\begin{center}
  \includegraphics[width=9 cm]{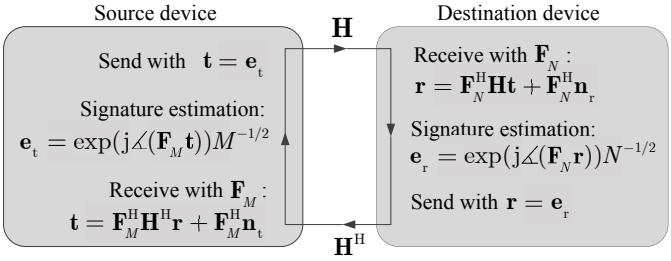}
  \caption{Process of the proposed STV scheme.}
  \label{fig:STV}
\end{center}  \vspace{-0.2 in}
\end{figure}

It is clear that given full CSI, the AWVs steering along the \emph{strongest} MPC in both ends can be obtained. In practical MMWC, however, ${\bf{H}}$ is basically unavailable in both ends; thus we propose the joint iterative beamforming training process of STV, which is shown in Fig. \ref{fig:STV}, and the corresponding algorithm is described in Algorithm \ref{alg:STV}. The iteration number $\varepsilon$ depends on practical channel response. According to the simulation results in Section V, $\varepsilon=2$ or 3 can basically guarantee convergence.

It is noted that STV is tailored for MMWC devices with CA phased arrays based on SGV. Thus, STV and SGV have common features, e.g., both schemes need iteration. However, their mathematical fundamentals are different. SGV is to find the principal singular vectors of the channel matrix ${\bf{H}}$, which is optimal and applicable for arbitrary channels, while STV is to find the CA steering vectors of the \emph{strongest} MPC by exploiting the directional feature, which is sub-optimal and only feasible under MMWC channels. Thus, in each iteration, STV requires the signature estimation, which is to estimate the CA steering vector of the \emph{strongest} MPC. Meanwhile, in order to make STV feasible for CA phased arrays, it adopts the DFT matrices in transmitting and receiving training sequences, because the entries of them have a constant envelope.

\begin{algorithm}[t]\caption{The STV Scheme}\label{alg:STV}
\begin{algorithmic}
\STATE \textbf{1) Initialize:}
\begin{quote}
Pick an initial transmitting AWV ${\bf{t}}$ at the source device. This AWV may be chosen either randomly or with the approach specified in Section V.
\end{quote}

\STATE \textbf{2) Iteration:}
Iterate the following process $\varepsilon$ times, and then stop.

\begin{quote}
Keep sending with the same transmitting AWV ${\bf{t}}$ at the source device over $N$ slots. Meanwhile, use discrete Fourier Transform (DFT) matrix ${\bf{F}}_N$ as the receiving AWVs at the destination device, i.e., the $n$th column of ${\bf{F}}_N$ as the receiving AWV at the $n$th slot. Note that ${\bf{I}}_N$ cannot be used for the receiving AWVs here, due to its non-constant-envelope entries, but other unitary matrices with constant-envelope entries are feasible. Consequently, we receiving a vector
${\bf{r}} = {\bf{F}}_N^{\rm{H}}{\bf{Ht}} + {\bf{F}}_N^{\rm{H}}{{\bf{n}}_{\rm{r}}}$,
where ${\bf{n}}_{\rm{r}}$ is the noise vector. Estimate the signature ${\bf{e}}_{\rm{r}}$ as ${{\bf{e}}_{\rm{r}}} = \exp ( {{\rm{j}}\measuredangle({{\bf{F}}_N}{\bf{r}}) )/\sqrt{N}}$
 and assign ${\bf{e}}_{\rm{r}}$ to ${\bf{r}}$.

 Keep sending with the same transmitting AWV ${\bf{r}}$ at the destination device over $M$ slots. Meanwhile, use DFT matrix ${\bf{F}}_M$ as the receiving AWVs at the source device. Consequently, we receiving a new vector
${\bf{t}} = {\bf{F}}_M^{\rm{H}}{{\bf{H}}^{\rm{H}}}{\bf{r}} + {\bf{F}}_M^{\rm{H}}{{\bf{n}}_{\rm{t}}}$,
where ${\bf{n}}_{\rm{t}}$ is the noise vector. Estimate the signature ${\bf{e}}_{\rm{t}}$ as ${{\bf{e}}_{\rm{t}}} = \exp ( {{\rm{j}}\measuredangle({{\bf{F}}_M}{\bf{t}}) )/\sqrt{M}}$ and assign ${\bf{e}}_{\rm{t}}$ to ${\bf{t}}$.
\end{quote}

\STATE \textbf{3) Result:}
\begin{quote}
${\bf{t}}$ is the AWV at the source device, and ${\bf{r}}$ is the AWV at the destination device.
\end{quote}
\end{algorithmic}
\end{algorithm}

\vspace{-0.0 in}
\section{Performance Evaluation}

\begin{figure*}
\begin{minipage}[t]{0.33\linewidth}
\centering
\includegraphics[width=6.2 cm]{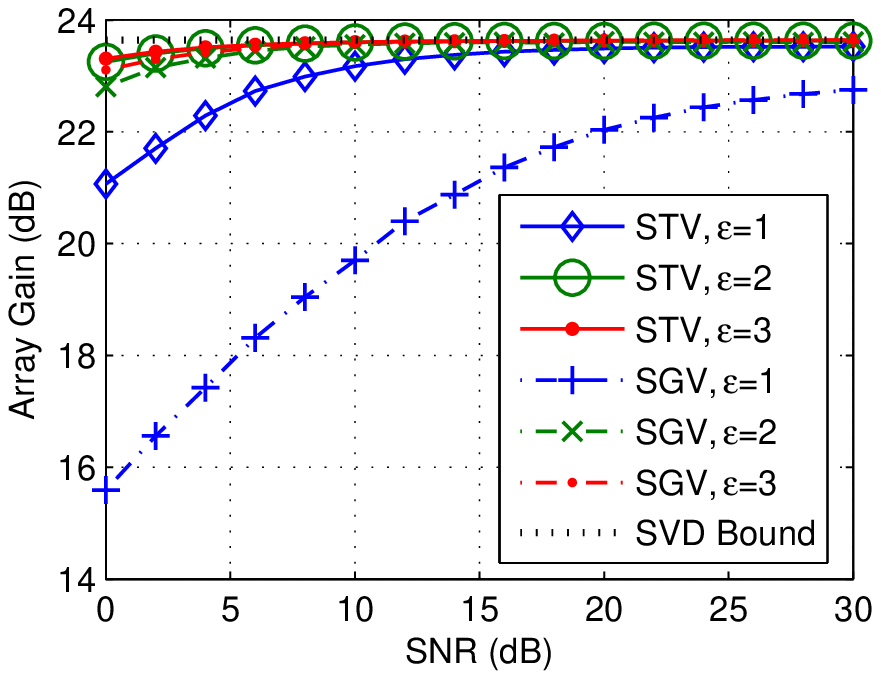}
\end{minipage} \hfill
\begin{minipage}[t]{0.33\linewidth}
\centering
\includegraphics[width=6.2 cm]{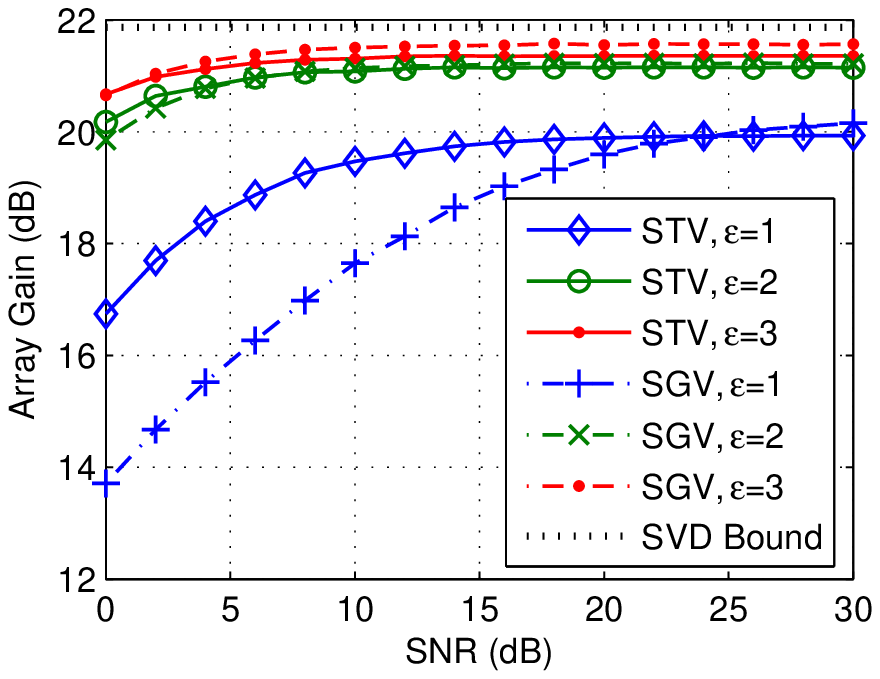}
\end{minipage}\hfill
\begin{minipage}[t]{0.33\linewidth}
\centering
\includegraphics[width=6.2 cm]{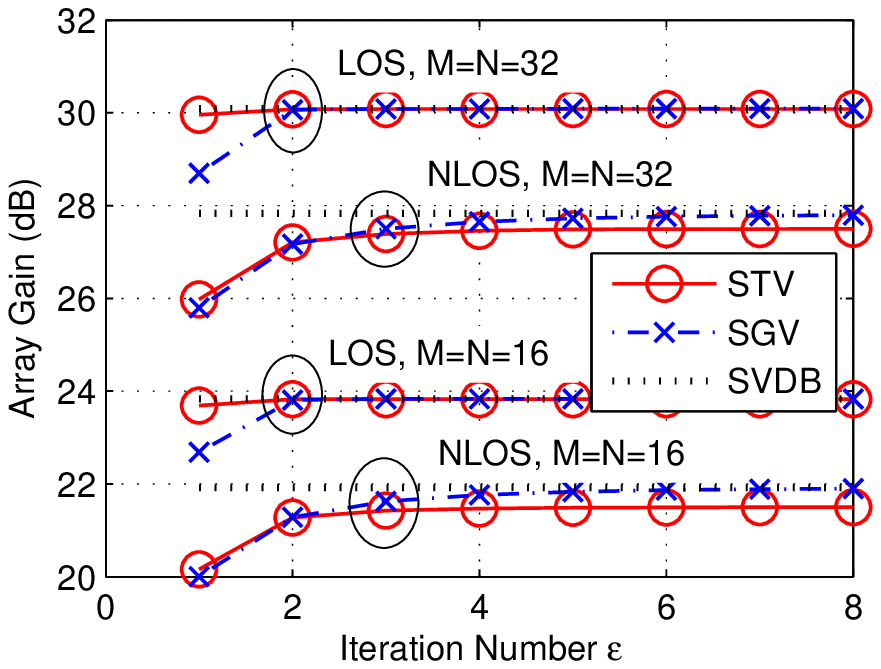}
\end{minipage}
\caption{Comparison of array gain between SGV and STV under the LOS (left) and NLOS (middle) channels with
different numbers of iterations, where $M=N=16$, and that of convergence rate (right), where SVDB represents SVD bound.}
\label{fig:cmp} \vspace{-0.1 in}
\end{figure*}

In this section we evaluate the performances of STV, including array gain and convergence rate, and compare them with those
of SGV via simulations.
In all the simulations, the channel is normalized as
$\mathbb{E}(\sum_{\ell=1}^L|\lambda_\ell|^2)=1$. The transmitting SNR is thus $\gamma_{\rm{t}}=1/\sigma^2$, and the array gain
becomes the ratio of receiving SNR to transmitting SNR, i.e.,  $\eta=\gamma/\gamma_{\rm{t}}=|{\bf{r}}^{\rm{H}}{\bf{H}}{\bf{t}}|^2$. The initial transmitting AWVs in the two schemes are selected under
the principle that its power evenly projects on
the $M$ basis
vectors of the receiving matrices at the source,
i.e., ${\bf{I}}_M$ and ${\bf{F}}_M$, respectively.
Thus, the initial transmitting AWV for SGV is ${\bf{1}}_M/\sqrt{M}$,
while that for STV is a normalized
constant-amplitude-zero-autocorrelation (CAZAC) sequence
with length $M$.

The array gain is empirically
found using the ratio of the average receiving SNR to
the average transmitting SNR over $10^3$ realizations of channels.
Furthermore, the SVD upper bound is obtained by averaging
the squares of the principal singular values of
these channel realizations.
Channel realizations are generated under
the Rician and Rayleigh fading models for
the LOS and NLOS channels, respectively.
For the LOS channel, the power of the LOS MPC is $|\lambda_1|^2=0.7692$,
and the average powers of the NLOS MPCs are
$\mathbb{E}(\{|\lambda_\ell|^2\}_{\ell=2,3,4})=[0.0769~0.0769~0.0769]$. For the NLOS channel, $\mathbb{E}(\{|\lambda_\ell|^2\}_{\ell=1,2,3,4})=[0.25~0.25~0.25~0.25]$. The transmitting and receiving steering angles are randomly generated within $[0~2\pi)$ in each realization.

The left and middle figures in Fig. \ref{fig:cmp} show the achieved array
gains of SGV and STV under the LOS and NLOS channels, respectively, with
different numbers of iterations, where $M=N=16$. The right figure in Fig. \ref{fig:cmp} shows the comparison of
convergence rate between SGV and STV under the LOS and NLOS channels
with a high transmitting SNR, i.e., 25 dB, in the cases of $M=N=16$ and $M=N=32$, respectively. From the left and right figures, it is found that under the LOS channel, both schemes achieve fast convergence
rates and approach the optimal array gain,
i.e., the SVD upper bound. From the middle and right figures, it is observed that under the NLOS channel, both
the two schemes have
slower convergence rates, and STV achieves a faster convergence rate at the cost of a slightly lower
array gain than SGV that also approaches the SVD upper bound. It is noted that, although not shown in these figures, similar results are observed with a smaller or larger number of antennas.

Explanations for these observations are as follows. Under the LOS channel,
there is one and only one strong MPC,
and the steering vectors of this MPC are almost the optimal AWVs.
Thus, STV can achieve the optimal array gain. But under the NLOS channel,
there are several MPCs with different steering angles (or steering vectors)
and the STV scheme obtains one of them as an AWV, which is not optimal.
Hence, STV cannot achieve the optimal array gain in such a case.
On the other hand, since the SGV is based on the principal singular vector,
it can surely achieve the SVD upper bound once convergence has
been achieved. Besides, the fact that STV achieves a faster convergence rate in NLOS channel indicates that the signature estimation in each iteration
of STV is more robust against noise, while the AWV estimation of SGV
is more sensitive to noise.

In brief, although STV is tailored for
MMWC devices with CA phased arrays, where SGV is infeasible, it has comparable performances
to SGV in terms of the convergence rate and array gain,
under both the LOS and NLOS channels. On the other hand, it is noted that a single iteration consumes $M+N$ training slots, which may significantly degrade the system efficiency, especially when the number of antennas is large. Hence, even if there is no CA constraint, i.e., both phase and amplitude are adjustable and thus SGV is feasible, STV may still be favored in the case that the iteration number is constrained to be 1 or 2 to save training time, because it achieves a higher array gain according to the right figure of Fig. \ref{fig:cmp}.


\vspace{-0.1 in}
\section{Conclusions}
Since the existing SGV scheme cannot be used
in MMWC with CA phased arrays, the STV scheme has been proposed
in this study, which effectively
exploits the directional feature of MMWC channels.
Performance comparisons showed that under LOS channel, both
the schemes achieve fast convergence
rates and achieve the optimal array gain; under NLOS channel, STV achieves
a faster convergence rate at the cost of a slightly lower array gain than
SGV that can still approach the optimal array gain. In summary,
while the proposed STV scheme is well-suited to
MMWC with CA phased arrays, it has comparable performances
to SGV in terms of the convergence rate and array gain
under both the LOS and NLOS channels.

\vspace{-0.1 in}


\end{document}